# Optical properties of recent non-fullerene molecular acceptors for bulk heterojunction solar cells

Andrea Farina[1], Giuseppe M. Paternò[2], Francesco Scotognella[2,3]*

[1] *Istituto di Fotonica e Nanotecnologie, Consiglio Nazionale delle Ricerche, Piazza Leonardo da Vinci 32, 20133 Milano, Italy*

[2] *Center for Nano Science and Technology@PoliMi, Istituto Italiano di Tecnologia, Via Pascoli 70/3, 20133 Milano, Italy*

[3] *Dipartimento di Fisica, Politecnico di Milano, Piazza Leonardo da Vinci 32, 20133 Milano, Italy*

## Abstract

For many years the faith of organic photovoltaics has been linked to the one of fullerene, since fullerene has been considered the electron-acceptor of choice in bulk heterojunctions solar cells. In the last years, the number of molecules that can be very competitive in replacing fullerene has increased significantly. In this work, we study by means of different theoretical methods five molecules that have demonstrated to work effectively as acceptors in organic heterojunctions. We discuss the comparison of simulated absorption spectra with the experimental spectra.

## Introduction

In the last decades, fullerene-based materials have been the most used acceptors in organic bulk heterojunction solar cells, owing to their relatively high processability and to the delocalization of the lowest unoccupied molecular orbital (LUMO) across the entire three-dimensional surface of fullerene [1]. Since the first report on the use of polymer:fullerene heterojunction as photovoltaic material in 1992 [2], the use of fullerene derivatives has seen a stark increase [3–9]. On the other hand, some intrinsic limitations of fullerenes, such the generally weak optical absorption in the visible and its environmental instability [10], have promoted the research of new non-fullerene acceptors [11]. The outstanding endeavour in chasing reliable substitutes to fullerene is, for example, testified by three exhaustive review articles published in 2019 [11–13]. Remarkably, power conversion efficiencies (PCE) above 16% have been achieved for solar cells integrating the non-fullerene acceptor BTPTT-4F [14–16]. In this context, also semiconducting carbon nanotubes have been proposed as valid alternative to fullerenes, due to their excitonic behaviour and relatively high environmental stability [17–19].

In this work, we simulate by employing different calculation methods the absorption properties of non-fullerene molecules that hold great promises as efficient electron- acceptor systems in organic heterojunction solar cells. The study of the optical gap and of the different optical transitions in non-fullerene acceptors permit to corroborate and complement the experimental studies present in the literature and aims at a better understanding of the different electronic transitions in the studied molecules.

## Methods

We have sketched the molecule geometries with the Avogadro package [20]. We have optimized the ground state geometries and we have calculated the electronic transitions of the molecules with the package ORCA 4.2.1 [21].

*Density Functional Theory calculations (with B3LYP functional)*: We have used the B3LYP functional [22] in the framework of the density functional theory. We have employed the Ahlrichs split valence basis set [23] and the all-electron nonrelativistic basis set SVPalls1 [24,25]. Moreover, the calculation utilizes the Libint library [26] and the Libxc library [27,28].

*Density Functional Theory calculations (with BP functional)*: We employ ORCA 4.2.1 [21] for these calculations. The orbital basis set def2-SVP has been used [29] and the auxiliary basis set def2/J has been used [30]. Also in this case, the calculation utilizes the Libint library [26] and the Libxc library [27,28].

*Calculations with Zerner's Intermediate Neglect of Differential Overlap (ZINDO/S), Modified Neglect of Diatomic Overlap (MNDO), Parametric Method 3 (PM3) methods*: Also for these calculations we employ ORCA 4.2.1 [21]. The orbital basis set def2-SVP has been used [29]. Also in this case, the calculation utilizes the Libint library [26] and the Libxc library [27,28].

*Calculations with Hückel method*: For the Hückel we employ the Hulis package [31,32].

**Results and Discussion**

In Figure 1, we show the molecular structures of the five investigated molecules. The molecule with 3-ethylrhodanine (RH) attached to both ends of thiophene-flanked carbazole is the so-called Cz-RH [33]. A solar cell that includes a bulk heterojunction with Poly(3-hexylthiophene) and Cz-RH (P3HT-Cz-RH) is reported in literature with a power conversion efficiency (PCE) of 2.56% [34].

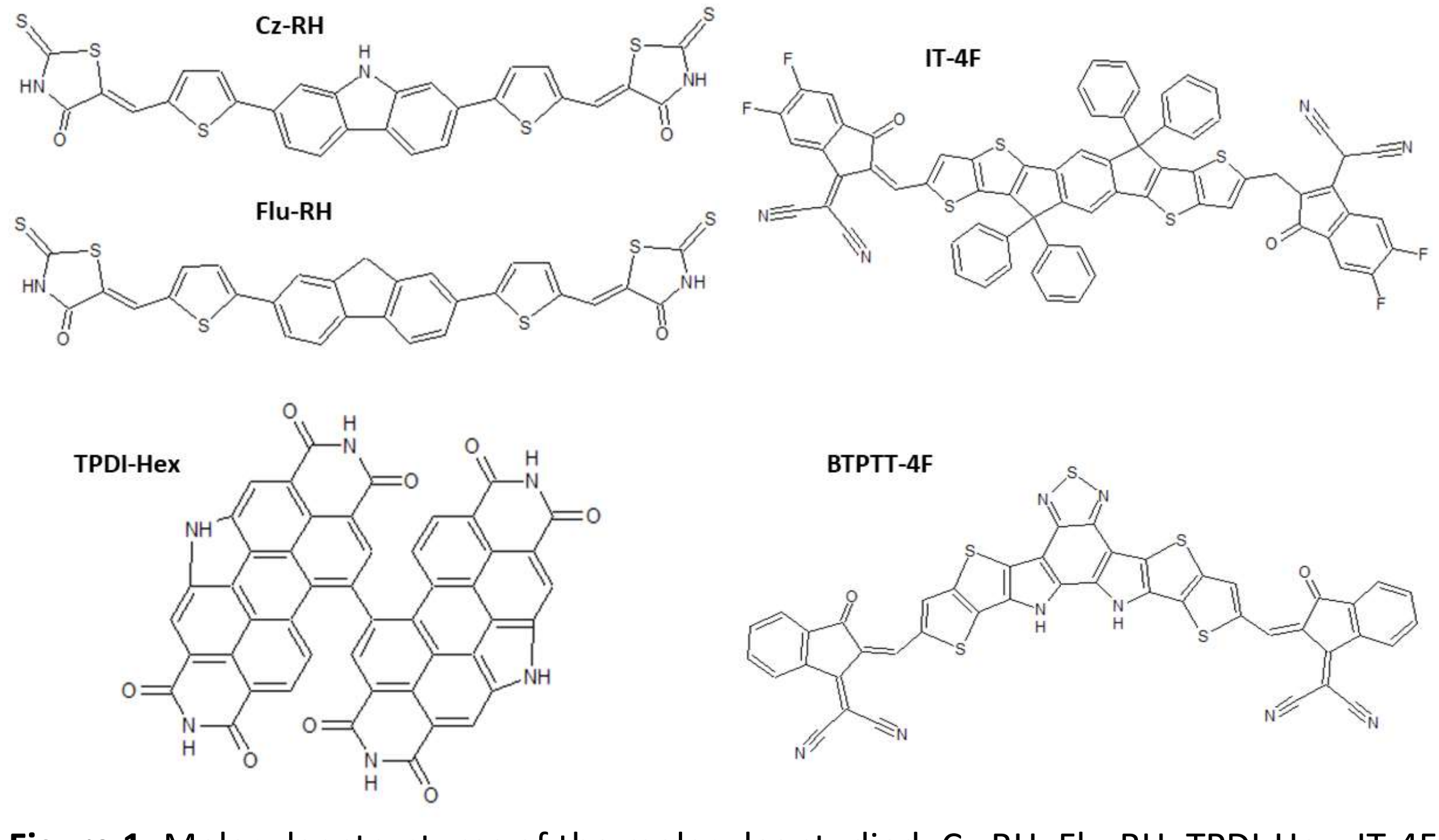

**Figure 1**. Molecular structures of the molecules studied: Cz-RH, Flu-RH, TPDI-Hex, IT-4F and BTPTT-4F.

The N-annulated perylene diimide (PDI) dimer has been employed in a bulk heterojunction solar cell reaching power conversion efficiency up to 7.6% with a terthiophene-based polymer named P3TEA as donor material [35]. The molecule IT-4F is used as acceptor in a bulk heterojunction with fluorinated poly[(2,6-(4,8-bis(5-(2-ethylhexyl)thiophen-2-yl)benzo[1,2-b:4,5-b′]dithiophene)-*co*-(1,3-di(5-thiophene-2-yl)-5,7-bis(2-ethylhexyl)benzo[1,2-c:4,5-c′]dithiophene-4,8-dione)] (PBDB-T-SF) leads to a PCE of 13% [36,37]. Finally, a heterojunction containing BTPTT-4F shows an efficiency above 16% as mentioned above [14].

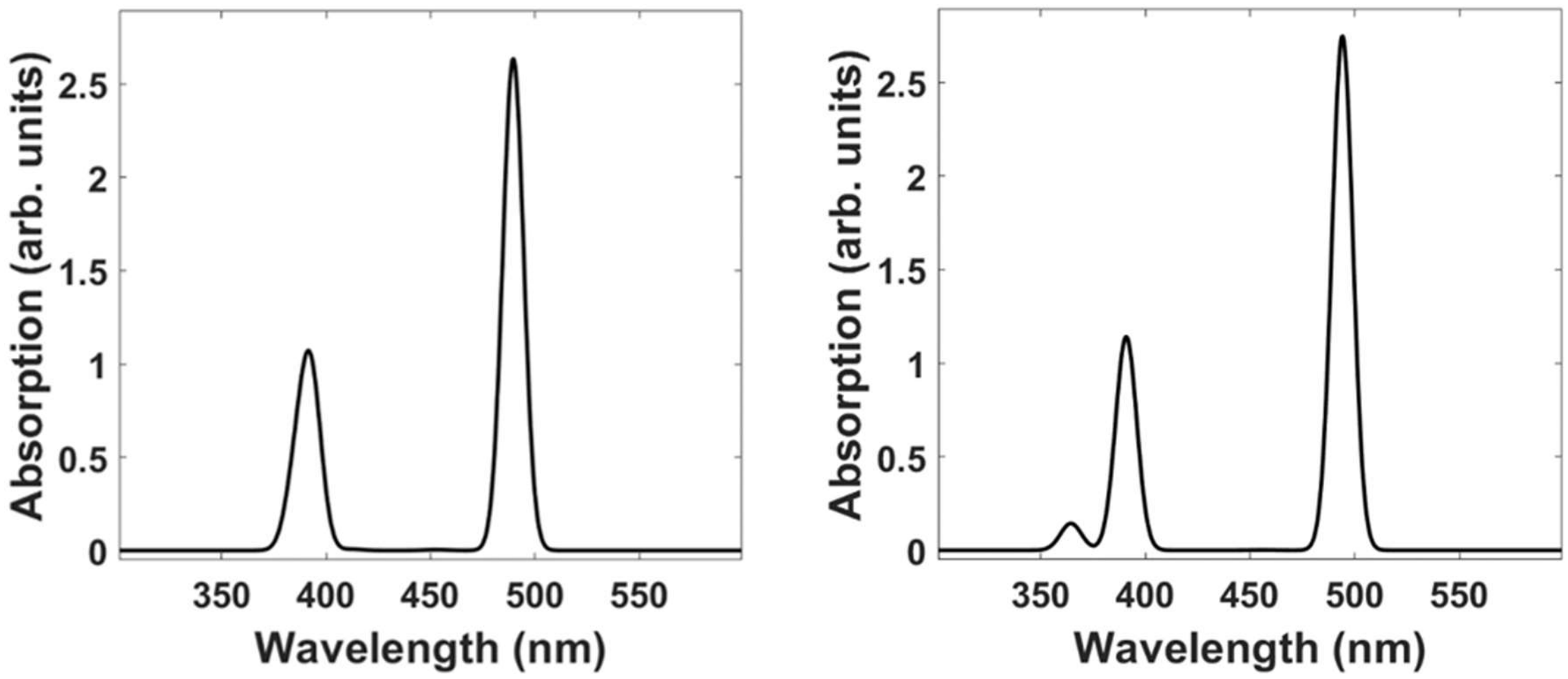

**Figure 2.** Calculation of the absorption spectrum (DFT with B3LYP functional) of the molecule Cz-RH (left) and the molecule Flu-RH (right).

We have calculated the first 16 transitions for the studied molecules (reported in the Supporting Information) and obtained simulated absorption spectra as a sum of Gaussian curves expressed as $f(x) = f_{osc}\exp[(x - x_c)^2/2a^2]$, with $f_{osc}$ the oscillator strength of the transition, $x_c$ the central wavelength (in nm) of the transition, $a$ that is related to the linewidth. In particular, we have selected a value of 5 nm for the linewidth $a$. In Figure 2, we show the simulations of the absorption spectra of Cz-RH molecule (left) and Flu-RH molecule (right). The lowest transition peak at around 500 nm (about 2.48 eV) is in good agreement with the experimental absorption spectra for the solutions reported in Kim at *al.* [34]. The highest predicted transitions are at longer wavelengths with respect to the experimental ones.

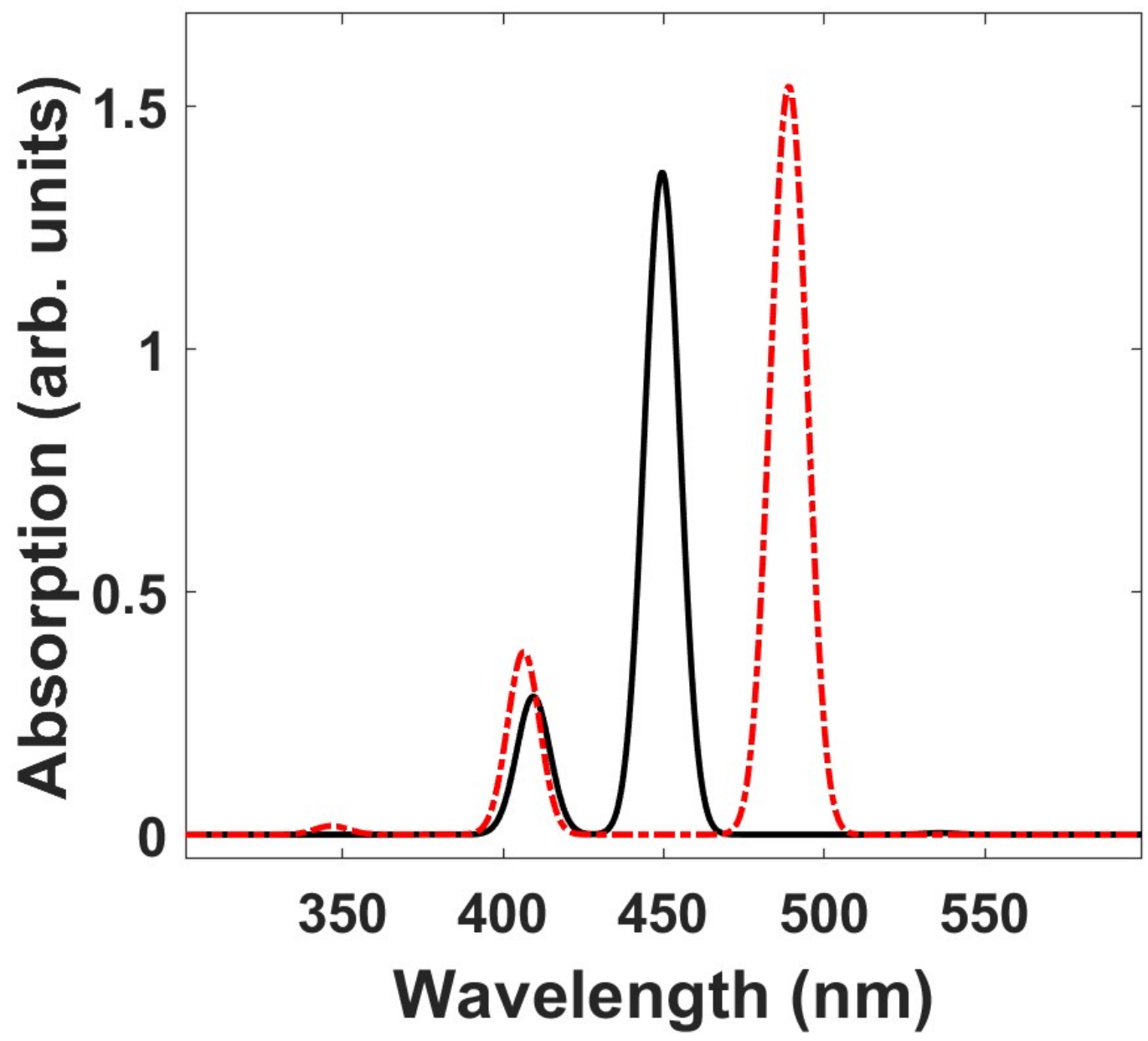

**Figure 3.** Calculation of the absorption spectrum of TPDI-Hex with DFT and B3LYP functional (solid black curve) and ZINDO/S method (dotted dashed red curve).

In Figure 3, we show the absorption spectrum of TPDI-Hex. We observe a discrepancy between the simulated absorption spectrum and the experimental one in terms of oscillator strength. The lowest simulated transition is at 2.31 eV (with a very weak oscillator strength of <0.01) while the lowest experimental transition is at 2.36 eV, but with a very strong weight in the spectrum with respect to the other peaks. With ZINDO/S method the simulated lowest transition is at 2.52 eV (with an oscillator strength of 0.93). These discrepancies could be due to the optimized geometry used and, in particular, to the dihedral angle between the perylene groups.

In Figure 4, we show the absorption spectrum of IT-4F with the lowest optical transition at 1.99 eV with DFT and B3LYP functional (black solid curve) and at 1.82 eV with ZINDO/S, while the experimental absorption shows the lowest absorption peak at 1.77 eV [36,37].

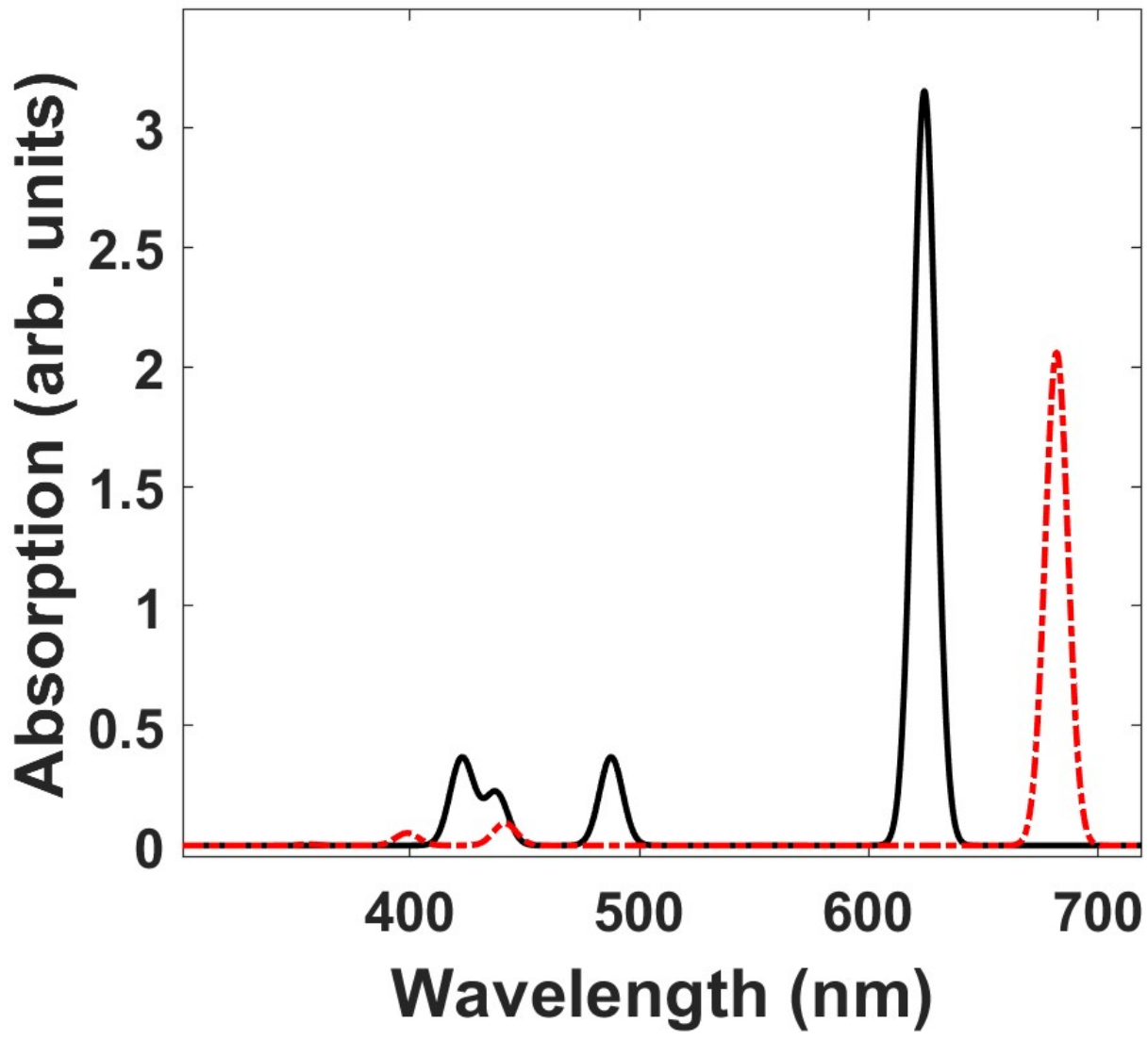

**Figure 4.** Calculation of the absorption spectrum of IT-4F with DFT and B3LYP functional (solid black curve) and with ZINDO/S method (dotted dashed red curve).

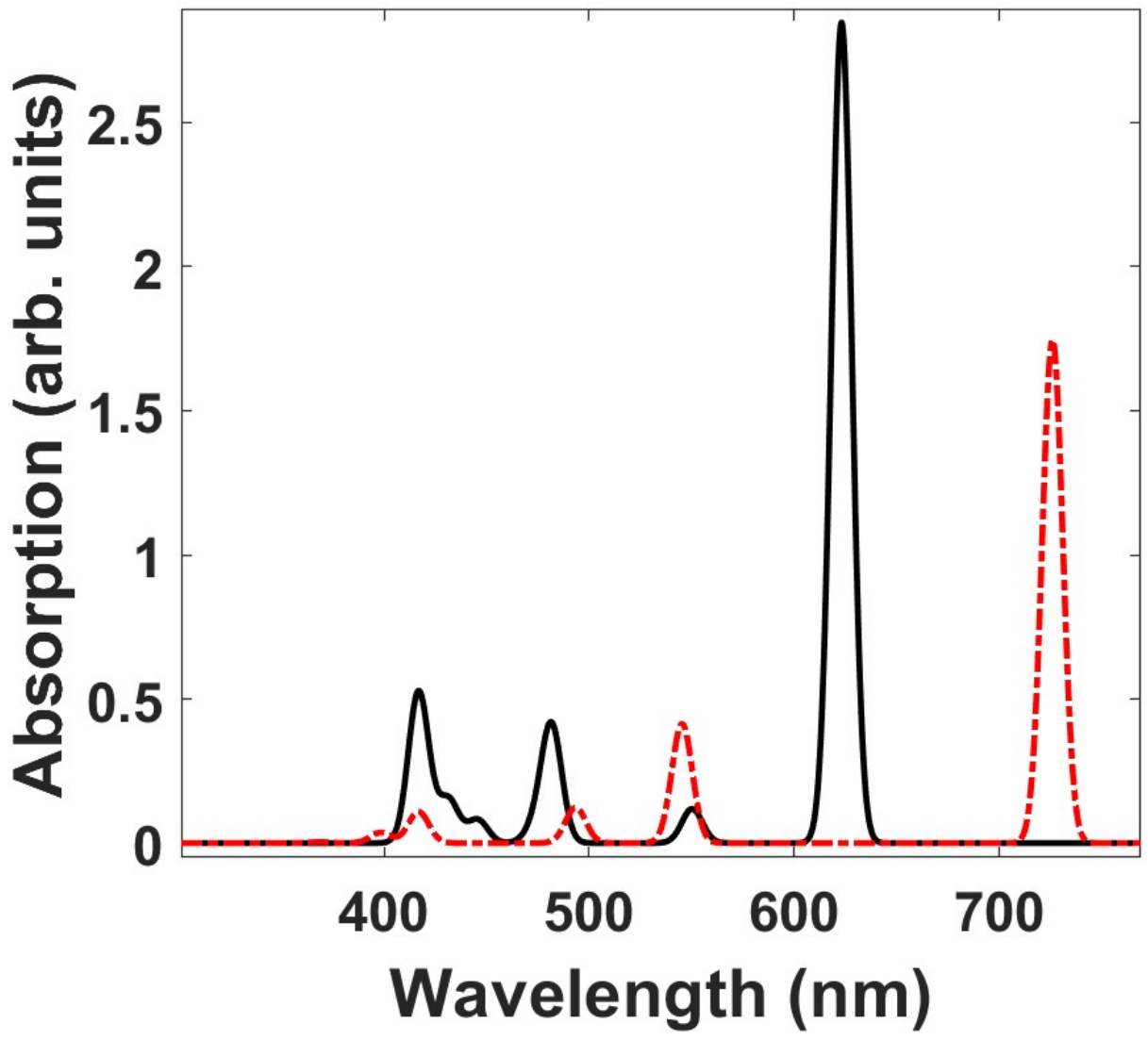

**Figure 5.** Calculation of the absorption spectrum of BTPTT-4F with DFT and B3LYP functional (solid black curve) and with ZINDO/S method (dotted dashed red curve).

In Figure 5, we display the absorption spectrum of BTPTT-4F, with the transition at 1.99 eV. For the estimation of BTPTT-4F optical bang gap, we have employed different theoretical methods whose results are reported in Table 1.

| | DFT B3LYP | DFT BP | ZINDO/S | MNDO | PM3 |
|---|---|---|---|---|---|
| Optical band gap (eV) | 1.99 | 1.65 | 1.71 | 2.01 | 2.28 |

**Table 1**. Optical band gap of BTPTT-4F calculated with different methods: DFT with B3LYP and BP functional, Zerner's Intermediate Neglect of Differential Overlap (ZINDO/S), Modified Neglect of Diatomic Overlap (MNDO), Parametric Method 3 (PM3).

Taking into account that Fan et al. [14] report an optical band gap of about 1.7 eV of BTPTT-4F solution
in chlorobenzene, DFT calculations with BP functional and ZINDO/S calculations give optical band gaps that are closer to the experimental one.

We perform Hückel method based calculations with Hulis package [31,32]. We find the following states: i) HOMO-1: $\alpha + 0.42\beta$; ii) HOMO: $\alpha + 0.24\beta$; iii) LUMO: $\alpha - 0.15\beta$; iv) LUMO+1: $\alpha - 0.16\beta$. Hence, we observe a HOMO-LUMO gap of $0.39\beta$. As reported by Fan et al. [14], the experimental HOMO, measured by cyclic voltammetry, is at -5.68 eV, while the experimental LUMO is at -4.06 eV. Thus, we could estimate a value of 4.15 for the $\beta$ parameter.

## Conclusion

In this work, we have studied the optical properties of five different non-fullerene acceptors by means of different calculation methods. These molecules, namely Cz-RH, Flu-RH, TPDI-Hex, IT-4F and BTPTT-4F, hold great promises for application in organic photovoltaic. In regards of BTPTT-4F, which has shown remarkable photovoltaic performances in organic heterojunction cells, we have determined the optical gap and compared it with the experimental results.

## Acknowledgement

G.M.P. thanks Fondazione Cariplo (grant n° 2018-0979) for financial support. This project has received funding from the European Research Council (ERC) under the European Union's Horizon 2020 research and innovation programme (grant agreement No. [816313]).

## Supporting Information

Cz-RH

ABSORPTION SPECTRUM VIA TRANSITION ELECTRIC DIPOLE MOMENTS

| State | Energy (cm-1) | Wavelength (nm) | fosc | T2 (au**2) | TX (au) | TY (au) | TZ (au) |
|---|---|---|---|---|---|---|---|
| 1 | 20425.0 | 489.6 | 2.636921809 | 42.50219 | -6.50477 | -0.06649 | 0.43090 |
| 2 | 21881.8 | 457.0 | 0.000280760 | 0.00422 | 0.05491 | -0.02345 | 0.02567 |
| 3 | 21884.8 | 456.9 | 0.000668999 | 0.01006 | -0.09583 | -0.01729 | 0.02410 |
| 4 | 22144.9 | 451.6 | 0.003095696 | 0.04602 | 0.01806 | 0.20002 | -0.07540 |
| 5 | 24264.7 | 412.1 | 0.006812285 | 0.09243 | 0.00579 | -0.30385 | -0.00816 |
| 6 | 25467.6 | 392.7 | 0.918517216 | 11.87340 | -3.43772 | -0.05104 | 0.23001 |
| 7 | 25912.9 | 385.9 | 0.117947044 | 1.49847 | 0.14975 | 1.21270 | -0.07351 |
| 8 | 25956.8 | 385.3 | 0.250613803 | 3.17855 | -1.77458 | 0.13092 | 0.11076 |
| 9 | 27101.0 | 369.0 | 0.000023860 | 0.00029 | -0.00361 | 0.01631 | 0.00330 |
| 10 | 27106.8 | 368.9 | 0.000033068 | 0.00040 | -0.00495 | 0.01900 | 0.00400 |
| 11 | 27614.2 | 362.1 | 0.151963043 | 1.81168 | 1.34274 | 0.01873 | -0.09157 |
| 12 | 29882.5 | 334.6 | 0.054487583 | 0.60028 | 0.00407 | -0.75792 | -0.16071 |
| 13 | 30997.0 | 322.6 | 0.026028282 | 0.27644 | -0.52413 | -0.00848 | 0.04072 |
| 14 | 31166.8 | 320.9 | 0.001844360 | 0.01948 | 0.00753 | 0.04215 | 0.13285 |
| 15 | 31499.9 | 317.5 | 0.000115857 | 0.00121 | -0.03455 | -0.00404 | 0.00079 |
| 16 | 31514.8 | 317.3 | 0.000127357 | 0.00133 | 0.03622 | -0.00223 | -0.00371 |

Flu-RH

ABSORPTION SPECTRUM VIA TRANSITION ELECTRIC DIPOLE MOMENTS

| State | Energy (cm-1) | Wavelength (nm) | fosc | T2 (au**2) | TX (au) | TY (au) | TZ (au) |
|---|---|---|---|---|---|---|---|
| 1 | 20238.5 | 494.1 | 2.748812476 | 44.71379 | 6.67292 | 0.05268 | -0.42800 |
| 2 | 21796.2 | 458.8 | 0.000098328 | 0.00149 | 0.00273 | 0.02174 | -0.03170 |
| 3 | 21797.2 | 458.8 | 0.000540438 | 0.00816 | 0.08993 | 0.00395 | -0.00769 |
| 4 | 22013.1 | 454.3 | 0.000930133 | 0.01391 | 0.00428 | 0.10743 | -0.04849 |
| 5 | 25578.7 | 391.0 | 1.073938741 | 13.82219 | 3.70979 | 0.06625 | -0.23501 |
| 6 | 25785.8 | 387.8 | 0.081275472 | 1.03766 | -0.15148 | 1.00541 | 0.06220 |
| 7 | 26944.6 | 371.1 | 0.000040679 | 0.00050 | -0.01780 | 0.01176 | 0.00648 |
| 8 | 26958.9 | 370.9 | 0.000027984 | 0.00034 | 0.01013 | -0.01400 | -0.00656 |
| 9 | 27443.0 | 364.4 | 0.143069407 | 1.71629 | -1.30724 | -0.01550 | 0.08467 |
| 10 | 29786.5 | 335.7 | 0.046371071 | 0.51251 | 0.00493 | 0.68594 | 0.20488 |
| 11 | 30940.8 | 323.2 | 0.015361888 | 0.16345 | -0.40366 | 0.00033 | 0.02257 |
| 12 | 31038.0 | 322.2 | 0.005879600 | 0.06236 | -0.00783 | -0.20036 | -0.14886 |
| 13 | 31359.3 | 318.9 | 0.000051734 | 0.00054 | -0.02286 | 0.00447 | 0.00083 |
| 14 | 31368.8 | 318.8 | 0.000115066 | 0.00121 | -0.03317 | -0.01019 | -0.00188 |
| 15 | 31484.3 | 317.6 | 0.000575788 | 0.00602 | 0.00268 | 0.04150 | 0.06551 |
| 16 | 31890.3 | 313.6 | 0.000009507 | 0.00010 | 0.00927 | 0.00098 | -0.00334 |

TPDI-Hex

DFT B3LYP

ABSORPTION SPECTRUM VIA TRANSITION ELECTRIC DIPOLE MOMENTS

| State | Energy (cm-1) | Wavelength (nm) | fosc | T2 (au**2) | TX (au) | TY (au) | TZ (au) |
|---|---|---|---|---|---|---|---|
| 1 | 18633.1 | 536.7 | 0.001188915 | 0.02101 | 0.07273 | -0.03149 | 0.12135 |
| 2 | 18666.3 | 535.7 | 0.002281881 | 0.04024 | 0.16884 | -0.08138 | 0.07151 |
| 3 | 22147.5 | 451.5 | 0.806393685 | 11.98665 | -1.99709 | -2.82305 | -0.16935 |
| 4 | 22367.8 | 447.1 | 0.694764420 | 10.22564 | 1.46986 | -0.89662 | -2.69466 |
| 5 | 24227.7 | 412.8 | 0.015520783 | 0.21090 | -0.11319 | -0.36717 | 0.25155 |
| 6 | 24257.6 | 412.2 | 0.015732386 | 0.21351 | -0.06679 | 0.45708 | 0.01135 |
| 7 | 24436.0 | 409.2 | 0.251680624 | 3.39075 | 1.10583 | -1.10701 | 0.97079 |
| 8 | 24754.8 | 404.0 | 0.012369662 | 0.16450 | 0.18705 | -0.09861 | -0.34611 |
| 9 | 26248.4 | 381.0 | 0.000181507 | 0.00228 | 0.00524 | 0.03785 | 0.02858 |
| 10 | 26264.2 | 380.7 | 0.000180823 | 0.00227 | 0.03578 | 0.01645 | -0.02675 |
| 11 | 26549.3 | 376.7 | 0.000097562 | 0.00121 | 0.01797 | -0.00552 | -0.02926 |
| 12 | 26550.5 | 376.6 | 0.000183039 | 0.00227 | 0.02252 | 0.04186 | 0.00320 |
| 13 | 29092.8 | 343.7 | 0.039767148 | 0.45000 | -0.38904 | 0.11121 | 0.53506 |
| 14 | 29099.1 | 343.7 | 0.040884198 | 0.46254 | 0.45870 | 0.45191 | 0.21888 |
| 15 | 29526.7 | 338.7 | 0.019212493 | 0.21421 | -0.03458 | 0.34812 | -0.30303 |
| 16 | 29540.0 | 338.5 | 0.032098441 | 0.35773 | 0.19759 | -0.55445 | 0.10613 |

ZINDO/S

ABSORPTION SPECTRUM VIA TRANSITION ELECTRIC DIPOLE MOMENTS

| State | Energy (cm-1) | Wavelength (nm) | fosc | T2 (au**2) | TX (au) | TY (au) | TZ (au) |
|---|---|---|---|---|---|---|---|
| 1 | 20353.5 | 491.3 | 0.929551124 | 15.03524 | 2.17953 | 3.20322 | 0.15590 |
| 2 | 20566.8 | 486.2 | 0.824032123 | 13.19025 | -1.68079 | 1.00422 | 3.05888 |
| 3 | 24606.3 | 406.4 | 0.370722871 | 4.95997 | -1.28858 | 1.39716 | -1.16081 |
| 4 | 24829.9 | 402.7 | 0.006722335 | 0.08913 | -0.12538 | 0.05557 | 0.26518 |
| 5 | 26308.3 | 380.1 | 0.000300105 | 0.00376 | -0.00872 | 0.05801 | -0.01772 |
| 6 | 26312.2 | 380.1 | 0.000141289 | 0.00177 | -0.01658 | -0.00055 | 0.03864 |
| 7 | 28382.8 | 352.3 | 0.001088332 | 0.01262 | 0.08222 | 0.07276 | 0.02387 |
| 8 | 28425.6 | 351.8 | 0.000536435 | 0.00621 | -0.03864 | 0.02199 | 0.06509 |
| 9 | 28883.2 | 346.2 | 0.017044440 | 0.19427 | 0.34113 | -0.15029 | 0.23519 |
| 10 | 28919.3 | 345.8 | 0.000728793 | 0.00830 | -0.02610 | 0.01597 | 0.08579 |
| 11 | 29147.3 | 343.1 | 0.085091643 | 0.96109 | -0.68880 | 0.45148 | -0.53180 |
| 12 | 29553.0 | 338.4 | 0.001389122 | 0.01547 | 0.05864 | -0.03093 | -0.10525 |
| 13 | 30740.3 | 325.3 | 0.001205058 | 0.01291 | -0.00910 | 0.07221 | 0.08722 |
| 14 | 30743.1 | 325.3 | 0.001194424 | 0.01279 | 0.08712 | 0.05875 | -0.04182 |
| 15 | 30873.6 | 323.9 | 0.000641028 | 0.00684 | 0.00639 | 0.05821 | 0.05836 |
| 16 | 30874.6 | 323.9 | 0.000634851 | 0.00677 | 0.06310 | 0.02686 | -0.04545 |

ITIC-4F

DFT B3LYP

ABSORPTION SPECTRUM VIA TRANSITION ELECTRIC DIPOLE MOMENTS

| State | Energy (cm-1) | Wavelength (nm) | fosc | T2 (au**2) | TX (au) | TY (au) | TZ (au) |
|---|---|---|---|---|---|---|---|
| 1 | 16021.3 | 624.2 | 3.151945843 | 64.76723 | -8.00573 | 0.81924 | 0.06586 |
| 2 | 18005.8 | 555.4 | 0.000078746 | 0.00144 | -0.00239 | -0.00050 | -0.03787 |
| 3 | 20505.4 | 487.7 | 0.369288338 | 5.92887 | -2.33422 | 0.69293 | 0.01248 |
| 4 | 20718.9 | 482.7 | 0.000018914 | 0.00030 | -0.00336 | 0.00127 | 0.01696 |
| 5 | 22754.0 | 439.5 | 0.000040494 | 0.00059 | 0.00134 | -0.00132 | -0.02413 |
| 6 | 22847.2 | 437.7 | 0.217244089 | 3.13034 | 1.74375 | -0.29914 | -0.01355 |
| 7 | 23430.8 | 426.8 | 0.063449449 | 0.89149 | -0.86504 | -0.37819 | 0.01301 |
| 8 | 23683.6 | 422.2 | 0.322638641 | 4.48482 | -2.11764 | -0.00381 | 0.02019 |
| 9 | 23723.4 | 421.5 | 0.000127641 | 0.00177 | -0.02349 | 0.00331 | 0.03477 |
| 10 | 24396.5 | 409.9 | 0.000517467 | 0.00698 | -0.00151 | -0.00026 | -0.08355 |
| 11 | 24933.1 | 401.1 | 0.045630591 | 0.60250 | -0.75371 | 0.18546 | 0.00547 |
| 12 | 25489.7 | 392.3 | 0.000001187 | 0.00002 | 0.00134 | -0.00031 | -0.00367 |
| 13 | 25745.6 | 388.4 | 0.001914986 | 0.02449 | -0.12496 | 0.09419 | -0.00059 |
| 14 | 25791.3 | 387.7 | 0.000225667 | 0.00288 | 0.00031 | 0.00177 | 0.05364 |
| 15 | 25927.0 | 385.7 | 0.000228345 | 0.00290 | -0.00019 | 0.00048 | 0.05384 |
| 16 | 26014.3 | 384.4 | 0.000883690 | 0.01118 | 0.05332 | 0.09129 | -0.00236 |

ZINDO/S

ABSORPTION SPECTRUM VIA TRANSITION ELECTRIC DIPOLE MOMENTS

| State | Energy (cm-1) | Wavelength (nm) | fosc | T2 (au**2) | TX (au) | TY (au) | TZ (au) |
|---|---|---|---|---|---|---|---|
| 1 | 14667.1 | 681.8 | 2.059597359 | 46.22895 | 6.76043 | -0.72285 | -0.05526 |
| 2 | 17736.9 | 563.8 | 0.000251870 | 0.00467 | -0.00038 | -0.00154 | -0.06836 |
| 3 | 22654.8 | 441.4 | 0.092704389 | 1.34715 | -0.31402 | -1.11721 | 0.01945 |
| 4 | 22761.7 | 439.3 | 0.000011792 | 0.00017 | -0.00030 | -0.00092 | 0.01302 |
| 5 | 25033.4 | 399.5 | 0.004722716 | 0.06211 | -0.21593 | 0.12443 | 0.00069 |
| 6 | 25037.1 | 399.4 | 0.000052211 | 0.00069 | -0.02218 | 0.01275 | 0.00566 |
| 7 | 25073.7 | 398.8 | 0.047222160 | 0.62002 | -0.68500 | 0.38830 | 0.00282 |
| 8 | 26636.3 | 375.4 | 0.000193395 | 0.00239 | 0.00054 | 0.00077 | 0.04888 |
| 9 | 28070.7 | 356.2 | 0.004324019 | 0.05071 | -0.00170 | -0.00283 | -0.22517 |
| 10 | 28173.3 | 354.9 | 0.005101163 | 0.05961 | -0.05964 | -0.23673 | 0.00293 |
| 11 | 29017.7 | 344.6 | 0.129519437 | 1.46943 | -0.36003 | 1.15743 | -0.01290 |
| 12 | 30150.3 | 331.7 | 0.527586376 | 5.76073 | -0.49875 | -2.34744 | 0.03893 |
| 13 | 30356.6 | 329.4 | 0.002217881 | 0.02405 | 0.00205 | 0.00364 | 0.15503 |
| 14 | 30995.4 | 322.6 | 0.001535878 | 0.01631 | -0.00075 | -0.00516 | -0.12762 |
| 15 | 31347.0 | 319.0 | 0.041230077 | 0.43301 | -0.09580 | 0.65098 | -0.00708 |
| 16 | 31662.5 | 315.8 | 0.000041130 | 0.00043 | -0.00086 | -0.00085 | -0.02064 |

BTPTT-4F

DFT B3LYP

ABSORPTION SPECTRUM VIA TRANSITION ELECTRIC DIPOLE MOMENTS

| State | Energy (cm-1) | Wavelength (nm) | fosc | T2 (au**2) | TX (au) | TY (au) | TZ (au) |
|---|---|---|---|---|---|---|---|
| 1 | 16047.9 | 623.1 | 2.847576407 | 58.41607 | 7.63172 | 0.41363 | -0.04208 |
| 2 | 18175.9 | 550.2 | 0.118290371 | 2.14254 | -0.09690 | 1.45952 | -0.05447 |
| 3 | 20765.2 | 481.6 | 0.406771754 | 6.44898 | -2.53572 | -0.13792 | 0.00920 |
| 4 | 21119.6 | 473.5 | 0.047485016 | 0.74020 | -0.04818 | 0.85786 | -0.04422 |
| 5 | 21257.5 | 470.4 | 0.002358955 | 0.03653 | 0.01275 | -0.19009 | 0.01532 |
| 6 | 22445.5 | 445.5 | 0.080785539 | 1.18490 | 1.08686 | 0.06001 | -0.00607 |
| 7 | 23185.9 | 431.3 | 0.147507340 | 2.09442 | -1.44469 | -0.08472 | 0.01019 |
| 8 | 23403.9 | 427.3 | 0.007233836 | 0.10175 | -0.31872 | -0.01283 | 0.00305 |
| 9 | 23991.1 | 416.8 | 0.525483349 | 7.21083 | 0.13973 | -2.68005 | 0.09282 |
| 10 | 24261.7 | 412.2 | 0.102003309 | 1.38410 | 1.17568 | 0.04285 | -0.00653 |
| 11 | 24834.9 | 402.7 | 0.151573874 | 2.00926 | 0.08111 | -1.41427 | 0.05031 |
| 12 | 25215.4 | 396.6 | 0.000805821 | 0.01052 | 0.01092 | 0.10165 | 0.00823 |
| 13 | 25245.7 | 396.1 | 0.000061430 | 0.00080 | 0.00551 | -0.02589 | -0.01003 |
| 14 | 26209.2 | 381.5 | 0.004901461 | 0.06157 | -0.01504 | 0.24705 | -0.01752 |
| 15 | 26493.3 | 377.5 | 0.012859732 | 0.15980 | 0.39871 | 0.02648 | -0.01127 |
| 16 | 27232.3 | 367.2 | 0.017631933 | 0.21315 | 0.01928 | -0.46086 | 0.01965 |

ZINDO/S

ABSORPTION SPECTRUM VIA TRANSITION ELECTRIC DIPOLE MOMENTS

| State | Energy (cm-1) | Wavelength (nm) | fosc | T2 (au**2) | TX (au) | TY (au) | TZ (au) |
|---|---|---|---|---|---|---|---|
| 1 | 13773.9 | 726.0 | 1.746736906 | 41.74901 | 6.45188 | 0.34797 | -0.03449 |
| 2 | 18337.0 | 545.3 | 0.413742478 | 7.42808 | -0.14737 | 2.71997 | -0.09002 |
| 3 | 20266.6 | 493.4 | 0.122671336 | 1.99268 | -1.40960 | -0.07498 | 0.01004 |
| 4 | 23979.0 | 417.0 | 0.044038567 | 0.60461 | 0.77742 | 0.01025 | 0.01096 |
| 5 | 24006.2 | 416.6 | 0.063920902 | 0.87659 | 0.02485 | -0.93318 | 0.07177 |
| 6 | 25109.7 | 398.3 | 0.036403897 | 0.47729 | 0.03797 | -0.68923 | 0.02837 |
| 7 | 25446.5 | 393.0 | 0.000007672 | 0.00010 | 0.00255 | -0.00322 | -0.00908 |
| 8 | 25452.7 | 392.9 | 0.000018754 | 0.00024 | -0.01057 | -0.00711 | -0.00896 |
| 9 | 27131.1 | 368.6 | 0.003406432 | 0.04133 | 0.20284 | 0.01141 | 0.00761 |
| 10 | 28539.6 | 350.4 | 0.150764843 | 1.73911 | 0.07150 | -1.31384 | 0.08848 |
| 11 | 29272.4 | 341.6 | 0.191892753 | 2.15812 | -1.46677 | -0.08173 | -0.00556 |
| 12 | 29721.5 | 336.5 | 0.010096900 | 0.11184 | -0.01540 | 0.33164 | -0.04019 |
| 13 | 30233.1 | 330.8 | 0.314421493 | 3.42378 | 1.84614 | 0.12341 | 0.01772 |
| 14 | 30421.1 | 328.7 | 0.495462123 | 5.36181 | -0.14231 | 2.30408 | -0.18103 |
| 15 | 31211.5 | 320.4 | 0.030617713 | 0.32295 | -0.03281 | 0.56559 | -0.04448 |
| 16 | 31768.1 | 314.8 | 0.004289706 | 0.04445 | 0.21045 | 0.01278 | -0.00103 |